\documentclass[conference,10pt]{IEEEtran}
\IEEEoverridecommandlockouts
\usepackage{cite}
\usepackage{amsmath,amssymb,amsfonts}
\usepackage{algorithmic}
\usepackage{graphicx, epstopdf}
\usepackage{textcomp}
\usepackage{xcolor}
\usepackage{url}            %
\usepackage{amsmath,amsthm,amsfonts,amssymb,amscd}
\usepackage[ruled,vlined,linesnumbered]{algorithm2e}
\usepackage{commath}
\usepackage{subcaption}
\usepackage{float}
\usepackage{multicol}
\usepackage{colortbl}
\usepackage{multirow, makecell}

\providecommand{\mT}{\mathcal{T} }

\begin{document}

\title{Penetrating RF Fingerprinting-based Authentication with a Generative Adversarial Attack\\
}

\author{\IEEEauthorblockN{Samurdhi Karunaratne, Enes Krijestorac, and Danijela Cabric 
\thanks{This work was supported in part by the CONIX Research Center, one of six centers in JUMP, a Semiconductor Research Corporation (SRC) program sponsored by DARPA.}
		}
		
\IEEEauthorblockA{\textit{Electrical and Computer Engineering Department,} \\
\textit{University of California, Los Angeles}\\
samurdhi@ucla.edu, enesk@ucla.edu, danijela@ee.ucla.edu }
}

\maketitle

\begin{abstract}
Physical layer authentication relies on detecting unique imperfections in signals transmitted by radio devices to isolate their fingerprint. Recently, deep learning-based authenticators have increasingly been proposed to classify devices using these fingerprints, as they achieve higher accuracies compared to traditional approaches. However, it has been shown in other domains that adding carefully crafted perturbations to legitimate inputs can fool such classifiers. This can undermine the security provided by the authenticator. Unlike adversarial attacks applied in other domains, an adversary has no control over the propagation environment. Therefore, to investigate the severity of this type of attack in wireless communications, we consider an unauthorized transmitter attempting to have its signals classified as authorized by a deep learning-based authenticator. We demonstrate a reinforcement learning-based attack where the impersonator---using only the authenticator's binary authentication decision---distorts its signals in order to penetrate the system. Extensive simulations and experiments on a software-defined radio testbed indicate that at appropriate channel conditions and bounded by a maximum distortion level, it is possible to fool the authenticator reliably at more than 90\% success rate.
\end{abstract}


\section{Introduction}

With the exponential growth of the Internet of Things (IoT), billions of new wireless devices are being deployed across the world every year \cite{statista2020}. The sheer number of devices available means that security systems that authenticate these devices should become cheaper, more secure and more robust. While traditional cryptography-based authentication systems have been the mainstay of wireless authentication, the unique requirements of IoT devices call for alternative methods that are sensitive to their computation and power constraints.

Passive Physical Layer Authentication (passive PLA) has been proposed as a low-overhead authentication method that requires little to no work on the part of the transmitter \cite{wang_wireless_2016}. Here, the authenticator uses channel state information and fingerprints due to hardware impairments to identify transmitters. Recently, research on passive PLA that uses deep learning techniques has been gaining momentum. Most such techniques process raw IQ samples from transmitters to extract features that are used to build classifiers. Since deep learning based classifiers tend to extract better features, these approaches have been shown to outperform others which use handcrafted features, reaching markedly higher accuracies \cite{riyaz_deep_2018}. However, there is an inherent vulnerability of deep learning-based classifiers to so-called \emph{adversarial examples}. For example, the existence of targeted adversarial examples have been pointed out \cite{szegedy2013}: given a valid input $x$, a classifier $C$ and a target $t$, it is possible to find an $x'=x+r$ such that $C(x') = t$ and $||r||$ is minimized. In this light, it is critical that deep learning-based PLA systems be analyzed for their vulnerabilities.

Radio fingerprints are usually considered hard to reproduce or replay because the replaying device suffers from its own impairments which disturb the features in the RF fingerprint. As such, naive replay attacks have limited success; only very recently has this problem been approached in more methodical ways. In \cite{shi2019generative}, Generative Adversarial Networks are used to train a spoofing device. While the method shows promising results, it relies on being able to place an adversarial receiver near the authenticating receiver, and only considers the case when there is one authorized transmitter. Furthermore, it is only verified through simulations where little information is given about how transmitter fingerprints were simulated. In \cite{restuccia2020}, this problem was investigated in a variety of angles, considering targeted and untargeted adversarial attacks, where the adversary tries either to make signals from a target transmitter $T_i$ be recognized as another specified transmitter $T_j$ (targeted) or be recognized as any transmitter other than $T_i$ (untargeted). They showed that spoofing can be done at a high accuracy both when the full gradients and the activations of the classifier are known; and only the activations of the final layer are known.  We identify the availability of both these as being practically unrealistic---the most we can expect from the authenticator is a binary feedback such as ACK or NACK denoting its decision.  Although such 1-bit feedback is technically compatible with their approach, the efficacy of the proposed method was not not evaluated in that regard. Additionally, their approach relies on signals from authorized transmitters being available to the adversary, and was only verified through training on offline data.

Inspired by this past work, we explore an adversarial attack that at most expects a binary feedback from the authenticator; is able to achieve high fooling rates in realistic channel conditions under a wide range of signal-to-noise ratios (SNRs); and attacks in real-time through online training. In this paper, we formulate this problem as a reinforcement learning problem and propose the use of policy gradient methods to perform spoofing of transmitters in a wireless network. Our results show that by distorting IQ samples of an adversarial transmitter---constrained to a maximum distortion level---before transmission, it is possible to fool a deep-learning based authenticator with high success rates even at low SNR and even when the only information available about the authenticator is a binary feedback received from it.

\section{System Model}

\begin{figure}[t]
    \begin{center}
    \vspace*{18px}
    \includegraphics[width=0.49\textwidth]{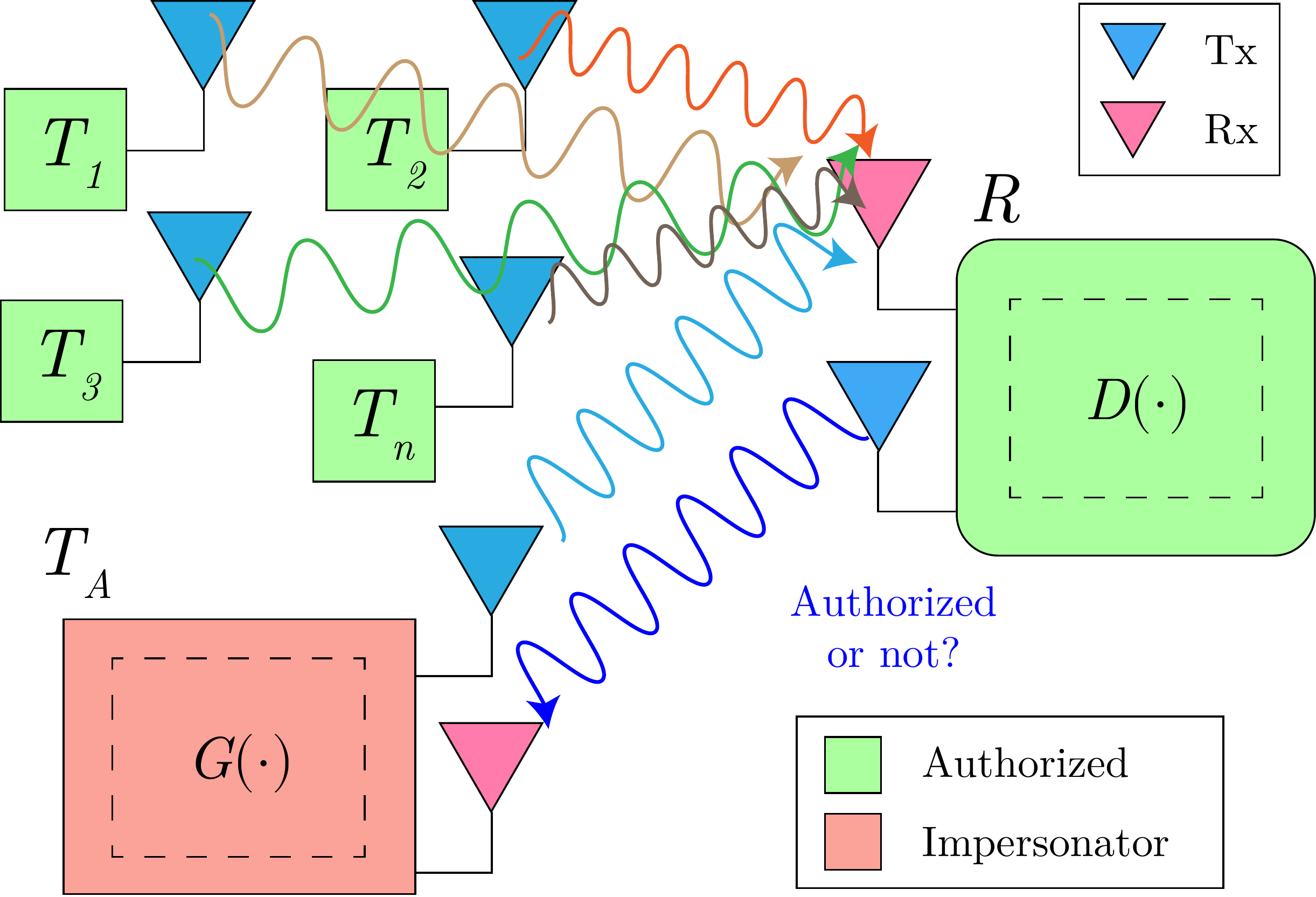}
    \end{center}
    \caption{System model considered in this project}%
    \label{fig:system_model}%
\end{figure}

We consider a wireless environment in which there are $|\mT|$ transmitters $\mT=\{T_1, T_2, \dots, T_{|\mT|}\}$ which are authorized to transmit to a single receiver $R$. $R$ is equipped with a pre-trained neural network-based authenticator $D$ that uses raw IQ samples of the received signals to perform a binary authentication decision at the physical layer, denoting whether the signal under consideration is from an authorized transmitter or not. There is an adversarial transmitter $T_A$ that wants to gain access to $R$, and it tries to do this by impersonating one of the authorized transmitters in $\mT$. $T_A$ employs a generator $G$ whose purpose is to distort the complex IQ samples of input discrete time signal $z(t)$ at $T_A$ such that after it is transmitted, it will be classified as authenticated at $D$. We assume that $D$ sends its authentication decision for each signal received from any transmitter, back to the transmitter. $T_A$ is also aware of the modulation being used by transmitters in $\mT$. This is visualized in Fig. \ref{fig:system_model}.

In a wireless communication system, there are three main sources of non-linearities that are imparted on the intended transmitted signal: if $x(t)$ is the signal at the beginning of the transmitter chain, the signal at the end of the receiver chain will be of the form $y(t)=f_R(f_C(f_T(x(t))))$, where $f_R$, $f_C$ and $f_T$ are non-linearities introduced by the receiver hardware, channel and transmitter hardware respectively. Since the channel is variable, $f_C$ is not a good a fingerprint and since there is only one receiver ($R$), $f_R$ will effectively be invariable across all transmitters. Therefore $D$ will be discriminating transmitters based on $f_T$.

We will try to spoof one of $\mT$ by only using the feedback from $R$. If we assume that $T_A$ could be placed reasonably close to transmitters in $\mT$, $f_C$ could be assumed to have the same variability for $T_A$ as well as for $\mT$. Such an assumption is justifiable when transmitters in $\mT$ are close together and are reasonably far away from $R$. This ensures that $D$ will discriminate signals from both $T_A$ and $\mT$ based on $f_T$.

\section{Proposed Solution}

We model this problem as a Markov Decision Process (MDP). An MDP is characterized by an \emph{agent} and an \emph{environment} that interact at each discrete time step $t$, whereby the agent selects an action $a_t$ according to a policy $\pi$ that takes into account the environment’s state $s_t$. In response to $a_t$, the agent receives a numerical reward $r_t$ from the environment and transitions to the next state $s_{t+1}$ \cite{sutton_reinforcement_2018}. We model $G$, represented with a neural network parameterized by $\theta$ (weights and biases), as the policy $\pi$ of the agent. The action $a_t \in \mathbb{R}^2$ such that $a_t \sim G_{\theta}(s_t)$, is the distorted value of $s_t$, and $r_t$ is a binary feedback received from $D$, which is part of the environment. The state $s_t$ can be represented in a number of ways, and will be discussed later in this section.

Assume the agent has collected a trajectory $\tau$ of length $\Gamma$ defined as a sequence of states, actions, and rewards, $\{s_1, a_1, r_1, s_2, a_2, r_2,\dots, s_{\Gamma}\}$. Now, the goal is to tune the parameters $\theta$ of $G$ under the following optimization problem:
\begin{align}
    \begin{array}{l}
         \text{maximize}_{\theta}~\mathbb{E}_{\tau}[J\vert G_{\theta}]  \\
         \text{where}~J= \sum_{t=1}^{\Gamma} \gamma^t r_t(s_t,a_t,s_{t+1})
    \end{array}
    \label{eq:optimization_problem}
\end{align}

Here $J$ is a metric of the the policy's performance, which is simply the cumulative reward of $\tau$ discounted by factor $\gamma$. To solve this, we can use a policy gradient method: we repeatedly estimate the gradient of the expected value of $J$ with respect to $\theta$ and use that to update $\theta$. To estimate the gradients, we will use a \emph{score function gradient estimator}. 
With the introduction of a baseline $b(s)$ to reduce variance, an estimate $\hat{g}$ for $\nabla_{\theta}\mathbb{E}_{\tau}[J]$ is \cite{schulman}
\begin{align}
\resizebox{.49 \textwidth}{!} 
{
    $\nabla_{\theta}\mathbb{E}_{\tau}[J(\tau)]\approx\hat{g}=\sum_{t=1}^{\Gamma} \nabla_\theta \log G_{\theta}(a_t|s_t) \left( \sum_{t'=t}^{\Gamma} r_{t'}\gamma^{t'-t} - b(s_t) \right)$
    \label{eq:gradient}
}
\end{align}
Now, the policy update can be done with any gradient ascent algorithm (e.g. $\theta \leftarrow \theta+\epsilon \hat{g}$). This can be repeated for a number of iterations, with a trajectory collected for each iteration, until $G$ converges to a satisfactory state. This algorithm also allows for $b(s)$ to be trained along with $\theta$ \cite{schulman}.

There are practical considerations when designing $r_t$ and $s_t$. For example, depending on the main type of distortion that the impersonator tries to mimic, different definitions of the state $s_t$ can be used.
\begin{enumerate}
    \item The state is a vector $s_t = \left[\text{Re}\{z(t)\}, \text{Im}\{z(t)\}\right]$ containing the real and imaginary part of the most recent IQ sample of the signal $z(t)$. This is applicable when the distortion over each sample is independent of the other samples. For example, the distortion imparted by the power amplifier in the RF chain will have this property \cite{hanna2019deep}.
    \item $s_t$ is $\{\left[\text{Re}\{z(t)\}, \text{Im}\{z(t)\}\right], H_{t-1}\}$, where $H_{t-1}$ is the hidden state of $G(s_{t-1})$, when it is modeled as a recurrent neural network. This state can in theory apply to any type of non-linearity.
\end{enumerate}
Irrespective of the particular definition of $s_t$, $a_t \sim G(s_t)$ is expected to reflect the distorted value of the most recent IQ sample; hence, $a_t \in \mathbb{R}^2$.

Although a reward $r_t$ is required for each $a_t$, a \emph{signal} transmitted by $T_A$ is usually a sequence of $N_s$ complex symbols (effectively a $N_s \times 2$ real vector) and hence a reward cannot immediately be obtained for each $a_t (t<\Gamma)$. So we set $\Gamma=N_s$ (i.e. the trajectory length is $N_s$) and transmit $A=\{a_1[0]+ja_1[1],a_2[0]+ja_2[1],\dots,a_{\Gamma}[0]+ja_{\Gamma}[1]\}$ as the signal to get the feedback $\hat{D}(A)=D(f_R(f_C(f_T(A)))$. To estimate $r_t$ from $\hat{D}(A)$, a Monte-Carlo search can be performed from $s_t$ until $s_{\Gamma}$, using a roll-out policy $G_{\beta}$ \cite{yu2017seqgan}. 
Specifically, at time $t$, if we have $\{s_1, a_1, r_1, s_2, a_2, r_2,\dots, s_t, a_t, r_t\} \sim G_{\theta}$, the rest of the trajectory is sampled from $G_{\beta}$ as $\{s_{t+1}, a_{t+1}, r_{t+1},\dots, s_{\Gamma}, a_{\Gamma}, r_{\Gamma}\} \sim G_{\beta}$ to produce a simulated $A_t$. Then for any $1 \leq t \leq \Gamma$, the reward $r_t$ can be written as
\begin{align}
r_t=\begin{cases}
 \frac{1}{M} \sum_{m=1}^{M} \hat{D}(A_t^{m}) & t<\Gamma \\
 \hat{D}(A) & t=\Gamma\\
\end{cases}
\label{eq:r_t}
\end{align}
where for $t<\Gamma$, we have averaged over $M$ Monte-Carlo searches due to the stochasticity of $G_{\beta}$.  In this approach $G_{\beta}$ is periodically updated to be the same as $G_{\theta}$; however, considering the large number of MC searches expected to be run, we can use a faster (and maybe less accurate) function approximator for $G_{\beta}$ \cite{silver_mastering_2016}.

We now add several optimizations to the method proposed above. First, to encourage exploration, we introduce entropy-regularization \cite{ahmed2019}: the agent gets a bonus reward at each time step proportional to $H(G_{\theta}(\cdot|s_t))$, the entropy of the policy at that timestep. i.e. $J$ in \eqref{eq:optimization_problem} changes to 
\begin{align}
J= \sum_{t=1}^{\Gamma} \gamma^t \left[r_t(s_t,a_t,s_{t+1})+ \beta H(G_{\theta}(\cdot|s_t))\right]    
\end{align}
where $\beta$ is the entropy coefficient (higher $\beta \implies$ more exploration). Also in practice, the components of a symbol cannot be distorted arbitrarily, as the decodability of the signal at the receiver side must be ensured. To integrate this constraint, we impose an action space limitation (clipping):
\begin{align}
a_t = \min(s_t+\epsilon|s_t|, \max(G(s_t), s_t-\epsilon|s_t|))
\end{align}
This effectively means that the maximum distortion level allowed is $\epsilon\%$ relative to the input state. 

Now we present an algorithm summarizing the proposed method above, depicted as Algorithm \ref{alg:cooperative}. Note that $G_{\theta}$ is initially trained by using Mean Squared Error (MSE) loss on a set of signals $\mathcal{S}_A$ captured from $T_A$, such that at the beginning there is no distortion ($G_{\theta}$ acts as an autoencoder). $G_{\beta}$ is initialized to $G_{\theta}$ and updated to $G_{\theta}$ periodically, after training $G_{\theta}$ for $g_{\text{steps}}$  iterations. This process is repeated for $K$ steps.

\begin{algorithm}
\begin{small}
\SetAlgoLined
\SetKwInOut{Input}{Input}
\SetKwInOut{Output}{Output}
\SetKwInOut{Initialization}{Initialization}
\Input{Generator policy $G_{\theta}$ roll-out policy $G_{\beta}$, Discriminator $D$, set of signals $\mathcal{S}_A$ from $T_A$\;
}
\Output  {$G_{\theta}$\;}
Initialize $G_{\theta}, G_{\beta}$\;
Pretrain $G_{\theta}$ using MSE on $\mathcal{S}_A$\;
$\beta \leftarrow \theta$\;
\For {$i={1,\dots,K}$}
{
\For {$g \in [1,g_{\text{steps}}]$}
 {
Generate a sequence $\{s_1,a_1,s_2,a_2,\dots,s_{\Gamma},a_{\Gamma}\} \sim G_{\theta}$\;
\For {$t={1,\dots,\Gamma}$}
{
Calculate $r_t$ using \eqref{eq:r_t}\;
}
Calculate $\nabla_{\theta}\mathbb{E}_{\tau}[J]$ using \eqref{eq:gradient}\;
Update $G_{\theta}$ using gradient descent\;
 }
$\beta \leftarrow \theta$\;
}
\end{small}
\caption{Generative adversarial attack with a cooperative $R$}
\label{alg:cooperative}
\end{algorithm}

\section{Experimental Evaluation}

This section is divided into three sections: Section \ref{sec:setup} details the simulation environment and the hardware testbed used for the evaluation, as well as the choices for different parameters; Section \ref{sec:nn_arch} presents the neural network architectures used for $D$ and $G$; and Section \ref{sec:results} describes four experiments conducted and the results obtained.

\subsection{Setup and parameters}
\label{sec:setup}
The proposed method was first evaluated on a simulated wireless environment written in Python. Power amplifier non-linearities $f_T(\cdot)$ are modeled by the Volterra Series, $f_T(z_t)=z_t(1 + \psi_0|z_t|^2 + \psi_1|z_t|^4)$, where $\psi_0$ and $\psi_1$ are coefficients unique to each transmitter, generated to follow a non-linear curve. Every transmitted packet of data consisted of completely random bits. We use QPSK modulation and root-raised-cosine (RRC) pulse shaping with 0.2 excess bandwidth.

Two channel models were investigated. The first one is a simple additive-white-gaussian-noise (AWGN) channel. The second one is a dynamic channel, which includes a set of more realistic impairments including timing errors, frequency errors, fading, inter-symbol interference and noise. The timing error is simulated by interpolating the signal by a factor of 32, choosing a random offset, and then downsampling. The frequency error is obtained by multiplying the signal with a complex exponential whose frequency  is selected from a Gaussian distribution with zero mean and 1 kHz standard deviation. For fading and inter-symbol interference, a three tap channel was used along with a Rayleigh coefficient of scale 0.5.

The state definition 2 in  Section 3 was used ($s_t$ is $\{\left[\text{Re}\{z(t)\}, \text{Im}\{z(t)\}\right], H_{t-1}\}$). The discount factor $\gamma$ was set to 1 (undiscounted). The gradient ascent on $\theta$ was done with the Adam optimizer, with the default configuration provided in the Keras API for Tensorflow, except that the learning rate was annealed starting from 0.001 to ensure convergence. We do not use any baseline function to train the generator, as entropy regularization and clipping already allow us to train the generator successfully. $\Gamma=N_s=256$, $\beta=1000$ and values for SNR, $\epsilon$, and $|\mT|$ were changed in different tests.

Finally, to test our attack on real hardware, we created a testbed consisting of 8 Analog Devices ADALM Pluto Software Defined Radios (SDRs); for convenience of operation, all were connected to a single computer. 6 SDRs were designated as authorized transmitters, 1 as an unauthorized transmitter and the other as the receiver, as shown in Fig. \ref{fig:fr_sdr_setup}. The Python module \texttt{pyadi-iio} was used to interface with the SDRs.
\subsection{\label{sec:nn_arch}Neural Network Architectures of $G$, $D$}

\begin{figure}[!t]
\begin{minipage}{.25\textwidth}
\begin{figure}[H]
    \includegraphics[width=.8\linewidth]{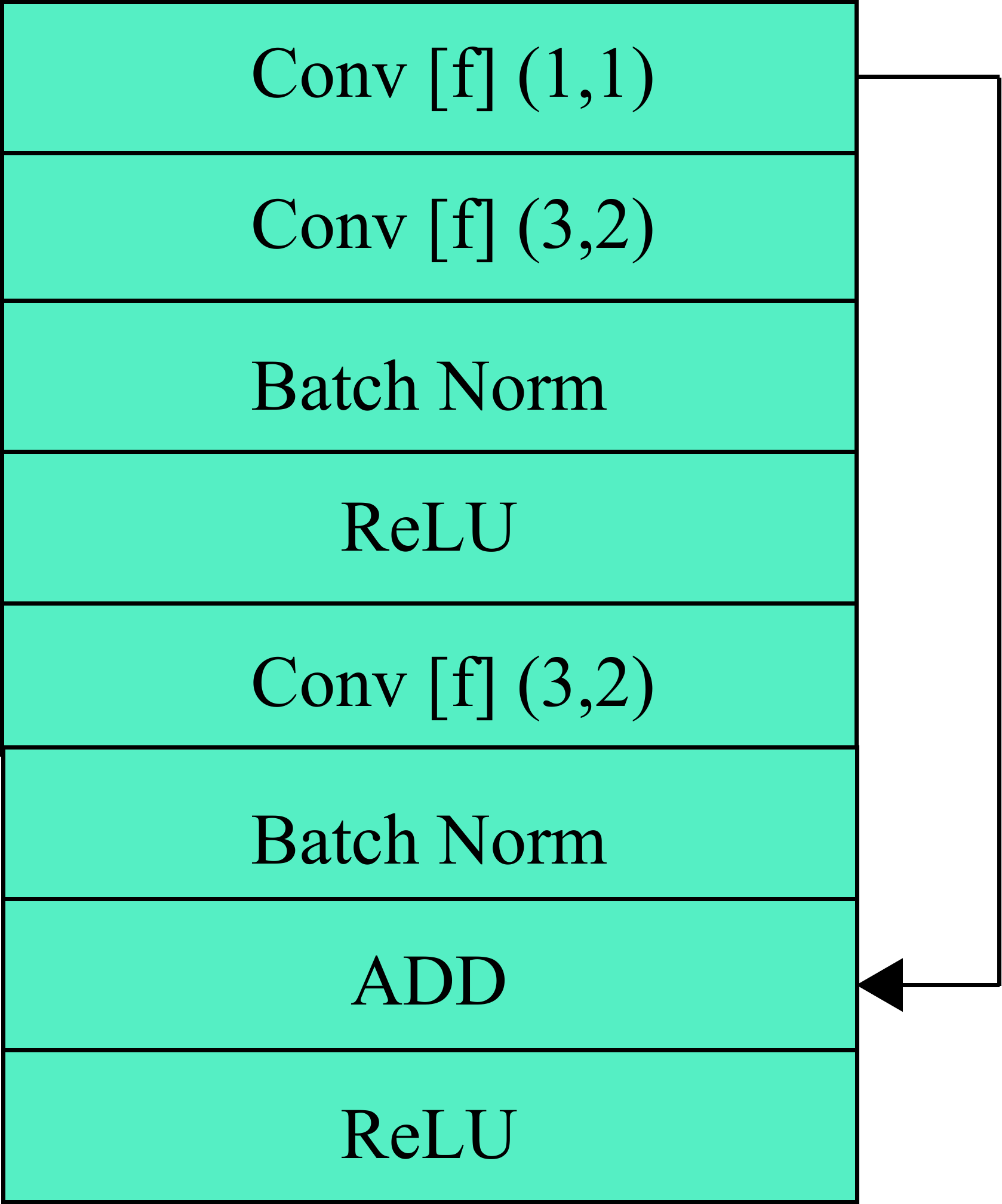}
    \caption{{Residual block $[f]$}}%
    \label{fig:residual_block}%
\end{figure}
\begin{figure}[H]
    \includegraphics[width=.8\linewidth]{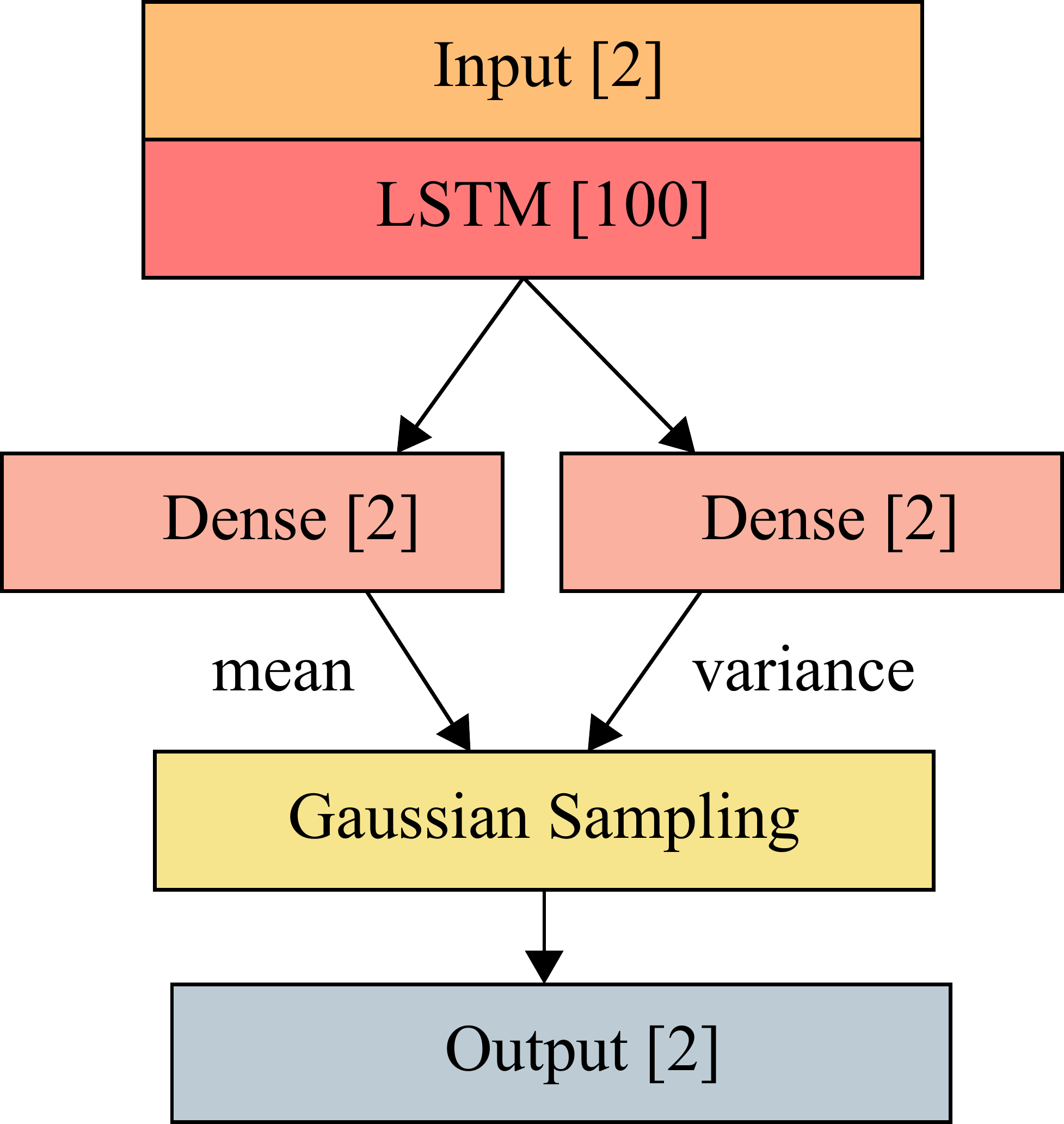}
    \caption{{Architecture of $G$}}%
    \label{fig:generator_arch}%
\end{figure}
\end{minipage}%
\begin{minipage}{.25\textwidth}
\begin{figure}[H]
    \includegraphics[width=.8\linewidth]{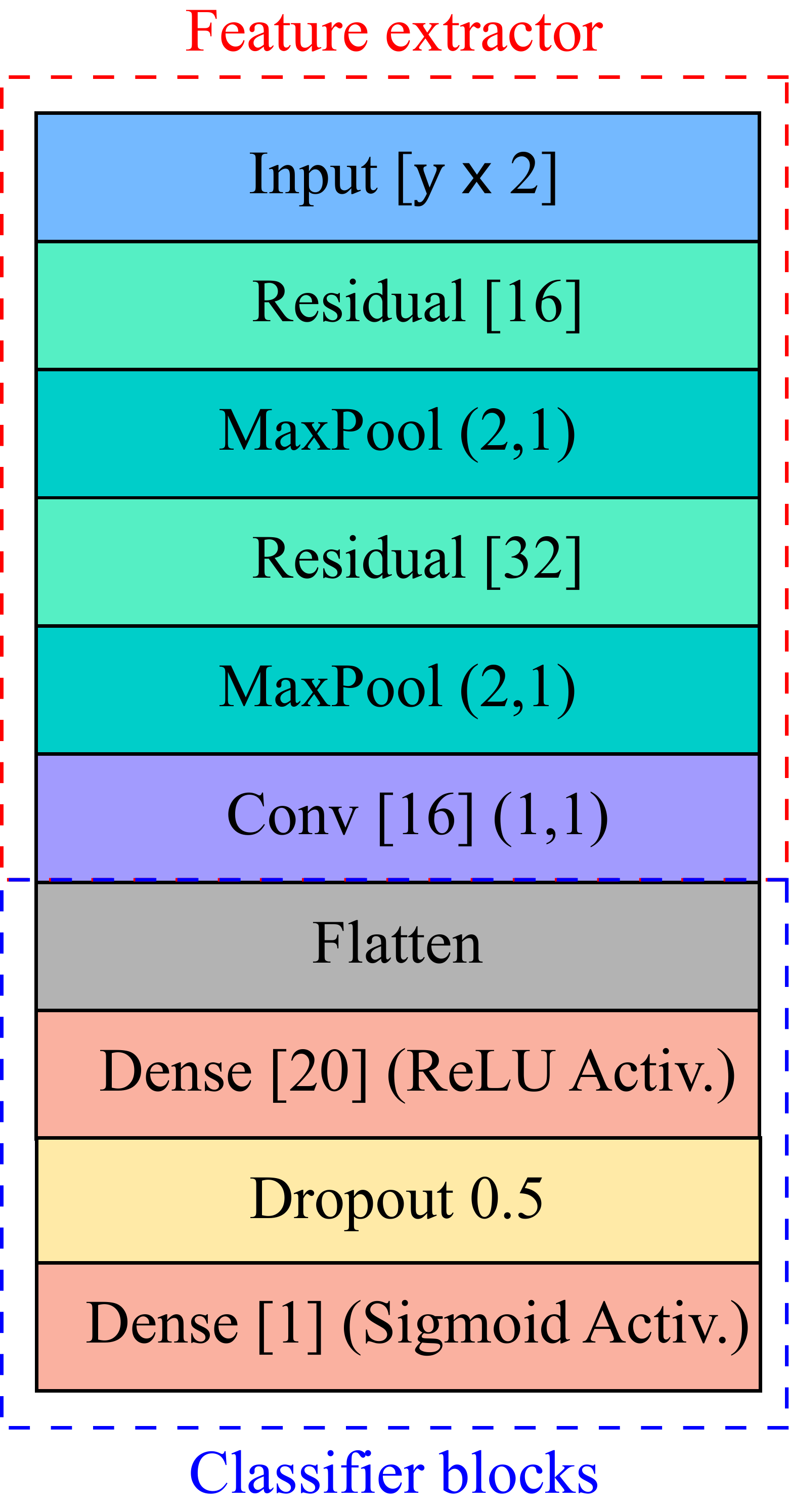}
    \caption{{Discriminator classifier $[y]$}}%
    \label{fig:discriminator_classifier}%
\end{figure}
\begin{figure}[H]
    \includegraphics[width=.8\linewidth]{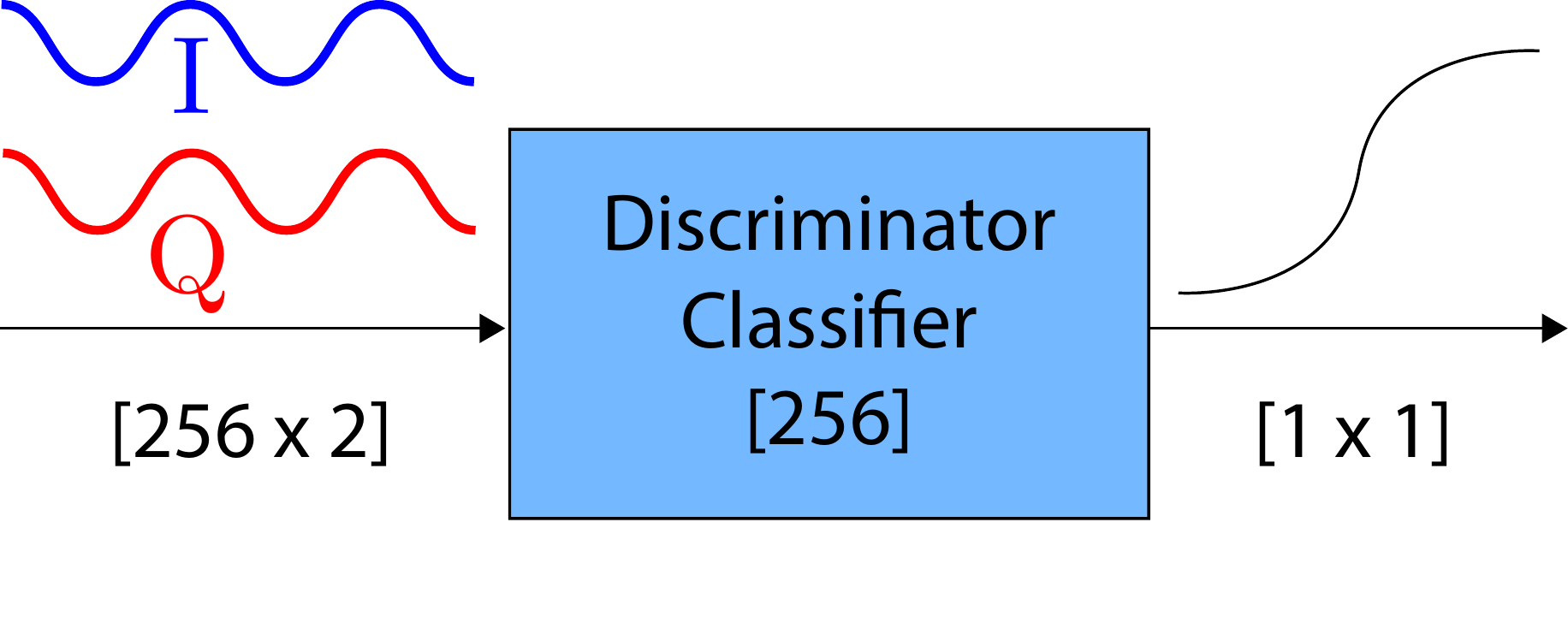}
    \caption{{Input for a simple channel}}%
    \label{fig:input_simple_channel}%
\end{figure}
\end{minipage}%
\begin{figure}[H]
    \begin{center}
    \includegraphics[width=1\linewidth]{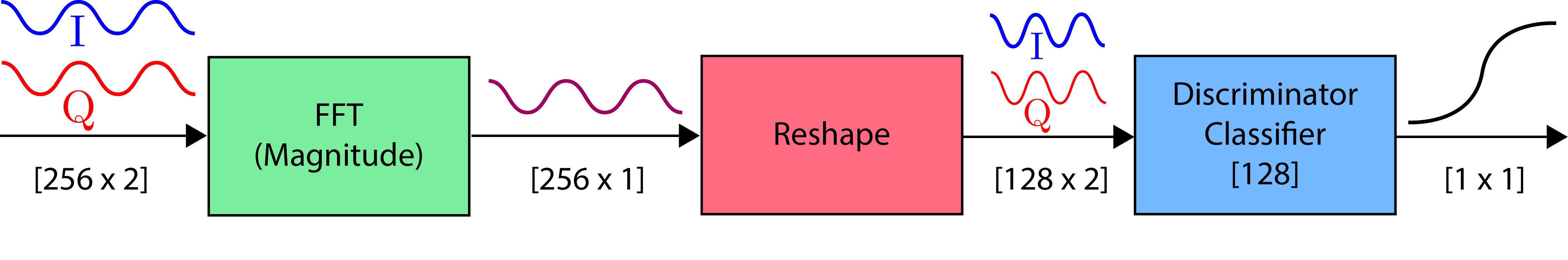}
    \end{center}
    \caption{{Input formats for discriminator architecture}}%
    \label{fig:input_dynamic_channel}%
\end{figure}
\end{figure}

We used a binary discriminator architecture for $D$, which has been shown to perform well in \cite{hanna_spawc_2020} for similar transmitter fingerprinting based classifications. It consists of a feature extractor consisting of a series of residual blocks with different numbers of filters, and a classifier block; the architecture of each type of block is shown in Fig. \ref{fig:discriminator_classifier}. Note that it produces a scalar output through a sigmoid activation; when providing binary feedback, this was thresholded at 0.5 to get a binary value (1 if greater than 0.5 and 0 otherwise). L2 regularization was used in the dense layers with weights of either 0.001 or 0.002 to avoid overfitting.

When using a simple channel, the IQ samples of the raw signal was passed to the discriminator classifier in Fig. \ref{fig:discriminator_classifier} without any pre-processing (each signal being a $(256 \times 2)$ vector). However, when using the dynamic channel model, this approach yielded poor discriminators with high fooling rates to begin with. So we first calculated the Discrete Fourier Transform of the raw signal, took the magnitude of the result, and reshaped it into a 2D signal of $(128 \times 2)$ before feeding to the discriminator. This pre-processing stage was chosen as it has been shown to produce superior results in similar transmitter fingerprinting based classifications \cite{hanna2019deep}. Fig. \ref{fig:input_simple_channel} and Fig. \ref{fig:input_dynamic_channel} depict this pictorially.

The architecture of $G_{\theta}$ (and $G_{\beta}$) is shown in Fig. \ref{fig:generator_arch}. Following the input, an LSTM layer with an output dimensionality of 100 was used, with the default configuration provided in Keras. Its outputs were modeled as the mean and the diagonal covariance of a two dimensional Gaussian distribution (one dimension each for the complex and imaginary part of the IQ samples), which was sampled to obtain the action, and to find the action probability (when calculating gradients) and entropy.

\subsection{Results}
\label{sec:results}
In this section, we report results of four experiments; Experiment 1-3 are conducted on the simulated environment and Experiment 4 is conducted on the hardware testbed.

\begin{figure}[t]
    \subfloat[{Fooling rate for different levels of SNR}]{\label{fig:fr_snr}\includegraphics[width=1\linewidth]{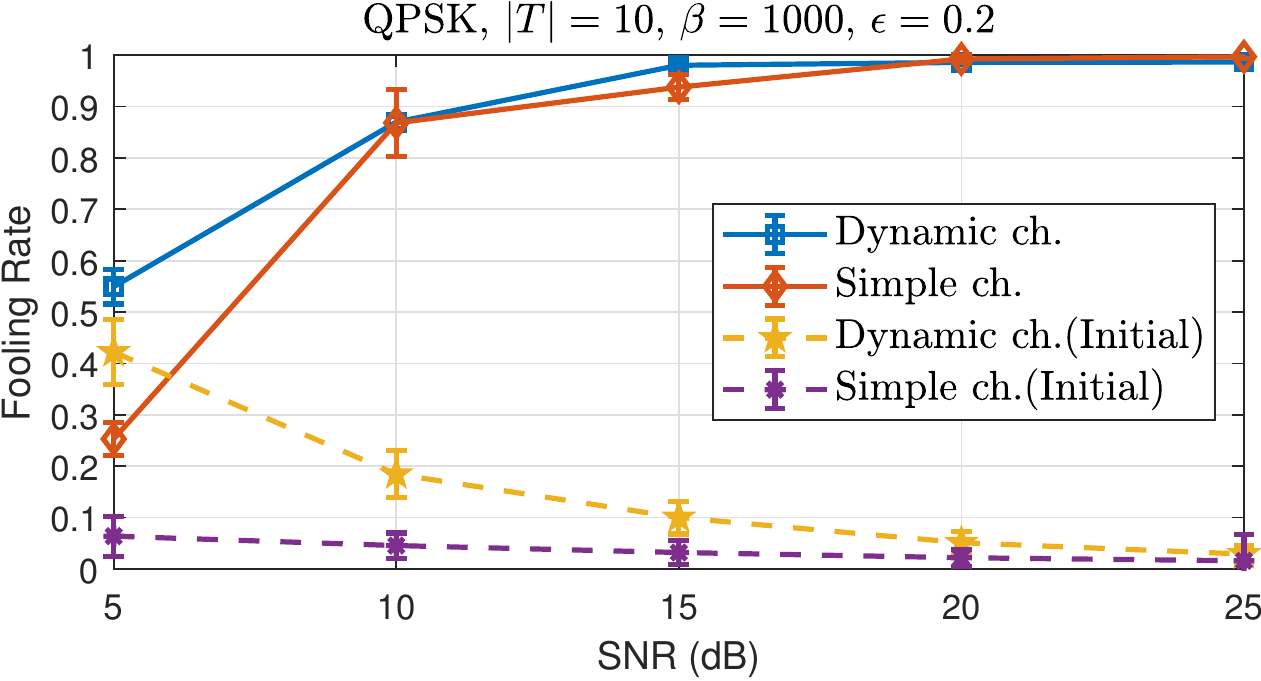}}
    
    \subfloat[{Convergence time for different levels of SNR}]{\label{fig:fr_steps}\includegraphics[width=1\linewidth]{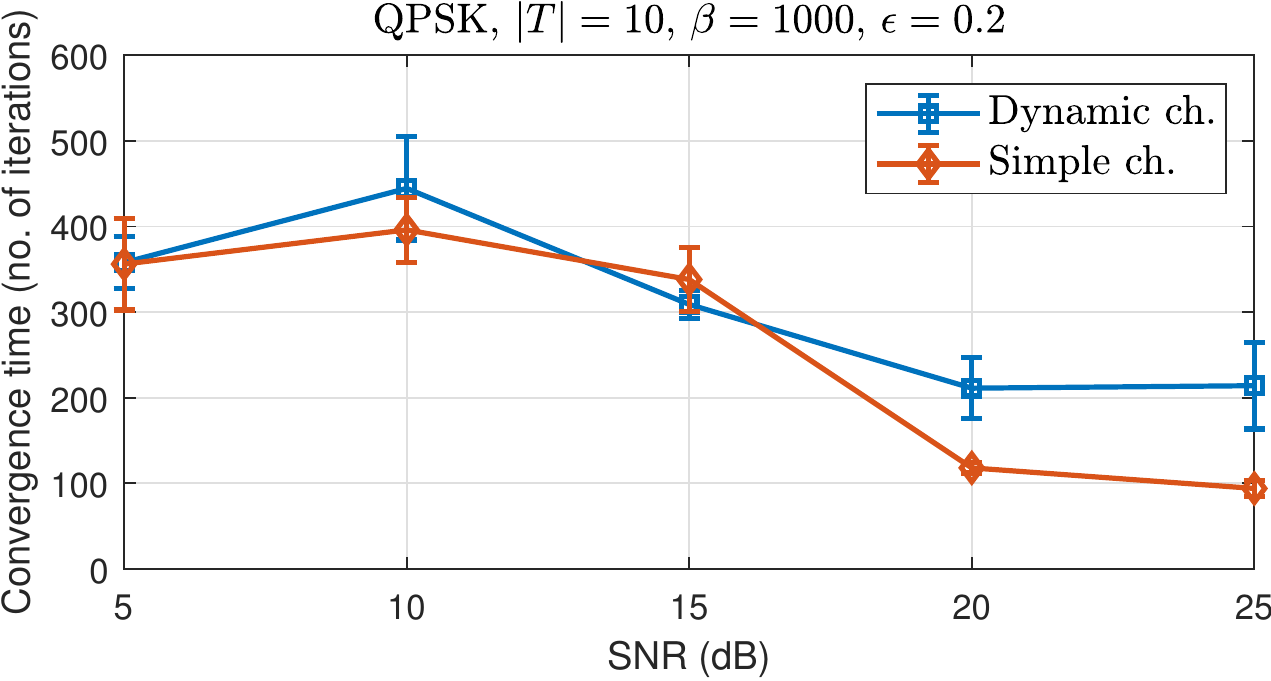}}
    \caption{{Fooling rate and corresponding convergence time for different levels of SNR for two channel models}}%
    \label{fig:fr_steps_snr}%
\end{figure}

For Experiment 1, we used a set of 10 authorized transmitters $|T|=10$ and a maximum distortion level of $\epsilon=0.2$. Then for five SNR values $\{5,10,15,20,25\}$ we evaluated the \emph{fooling rate} of $D$ at convergence for both channel models. The results are shown in Fig. \ref{fig:fr_snr}. The dashed lines show the initial fooling rate; for the simple channel, it starts at around 7\% for low SNR and decreases to near 0\% for higher SNRs. For the dynamic channel, the initial fooling rates are higher but still less than 10\% for even moderately high SNRs. This shows that the discriminator performs excellently at the beginning (except for the case of 5 dB SNR for the dynamic channel). It is clear that even at really low SNR, significant increases in fooling rate can be achieved, with near 100\% fooling rates being achieved at and above 20 dB SNR for both types of channels. Although slightly higher fooling rates are achieved for the dynamic channel at certain SNRs, this should be put in perspective with the higher initial fooling rates of the discriminator in the dynamic channel---in fact, the simple channel gives a higher relative improvement. Fig. \ref{fig:fr_steps} denotes the convergence time corresponding to Fig. \ref{fig:fr_snr}, measured by the number of gradient descent updates of $G$ (number of iterations of the inner-loop of Algorithm \ref{alg:cooperative}). As expected, we see that algorithm converges faster for higher SNRs, except for the jump from 5 dB to 10 dB. This is due to the fooling rate gain at 5 dB being much smaller than at 10 dB, and hence the algorithm achieving that smaller gain in a less number of iterations.
For a practical perspective, consider the simple channel at 20 dB SNR with convergence time of roughly 100 iterations. Since each iteration requires 256 feedbacks, we need 25600 feedbacks in total from $D$. Although this might seem excessive, assuming the system environment stays fairly static, we can space out the attack say, over 24 hours to reduce suspicion. This means that we only need a feedback every $\approx 3.4~s$, which is several orders of magnitude larger than a typical packet transmission time (e.g. 1 ms)---if we do not desire a near 100\% fooling rate, this time-interval can be greatly increased.

\begin{figure}[t]
    \begin{center}
    \includegraphics[width=1\linewidth]{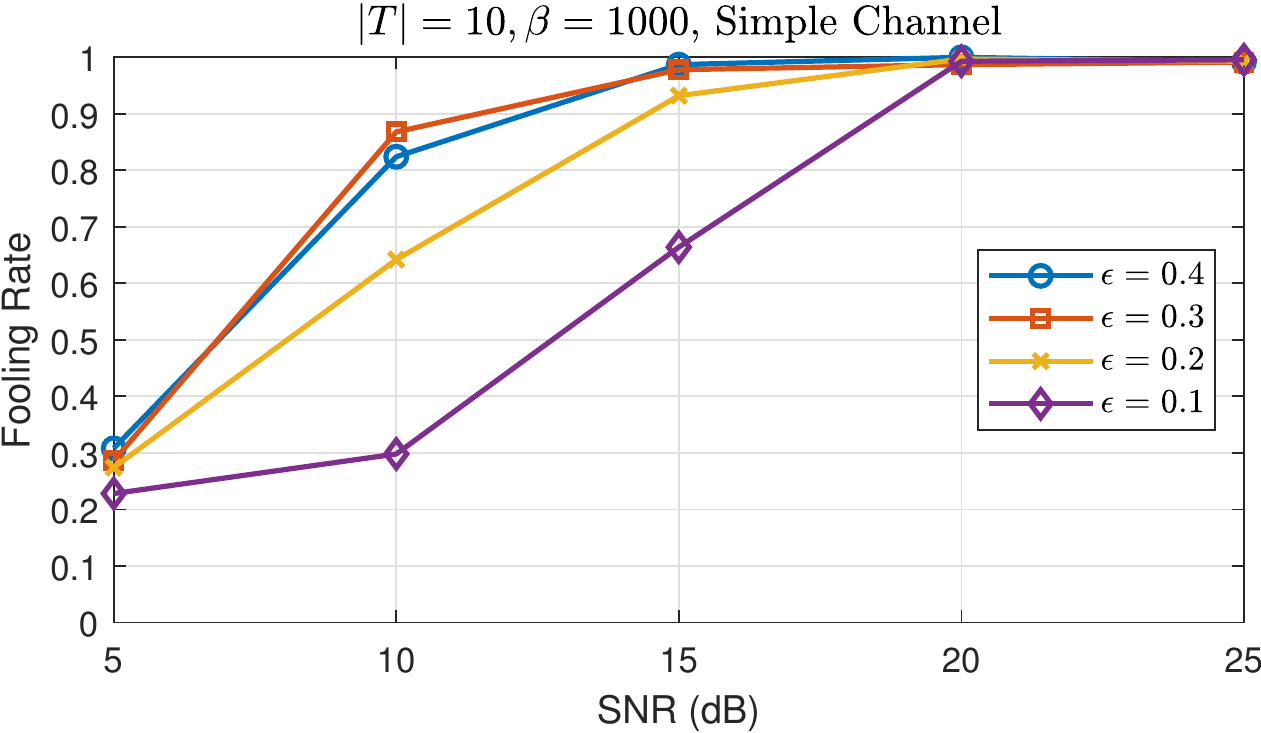}
    \end{center}
    \caption{{Effect of changing $\epsilon$ on the fooling rate at different SNRs}}%
    \label{fig:snr_eps}%
\end{figure}

In Experiment 2, we wish to evaluate the effect of the maximum distortion level $\epsilon$ on the fooling rate---specifically, we seek justification for our intuition that allowing $G$ more freedom for distortion should allow it to reach higher fooling rates. Fig. \ref{fig:snr_eps} shows the results obtained when the fooling rate was evaluated for $\epsilon \in \{0.1,0.2,0.3,0.4\}$, and for the same $\mT$ and $\beta$ as in Experiment 1, but only for the case of a simple channel. As expected, we see that a higher $\epsilon$ most certainly allows a higher fooling rate to be achieved and that a sufficiently high $\epsilon$ allows for near 100\% fooling rates. This means that by limiting $\epsilon$, we can still launch a successful attack, while keeping the amount of distortion imparted on the transmitted signals at a controlled level.

\begin{figure}[t]
    \centering
    \subfloat[{SDR testbed with $|\mT|=6$}]{\label{fig:fr_sdr_setup}\includegraphics[width=0.38\linewidth]{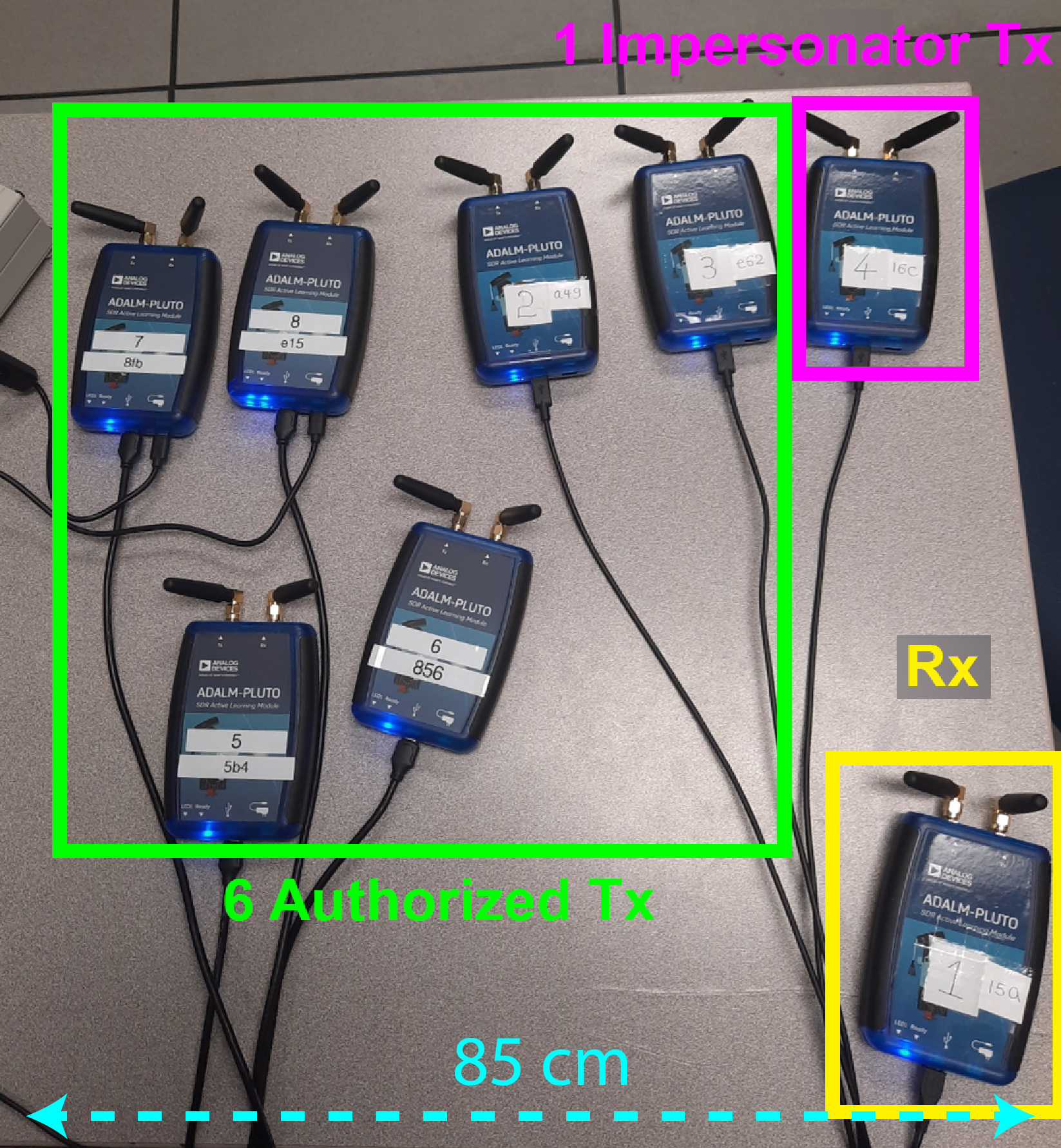}}\hfill
    \subfloat[{Effect of changing $\epsilon$ on the fooling rate}]{\label{fig:fr_sdr_results}\includegraphics[width=0.6\linewidth]{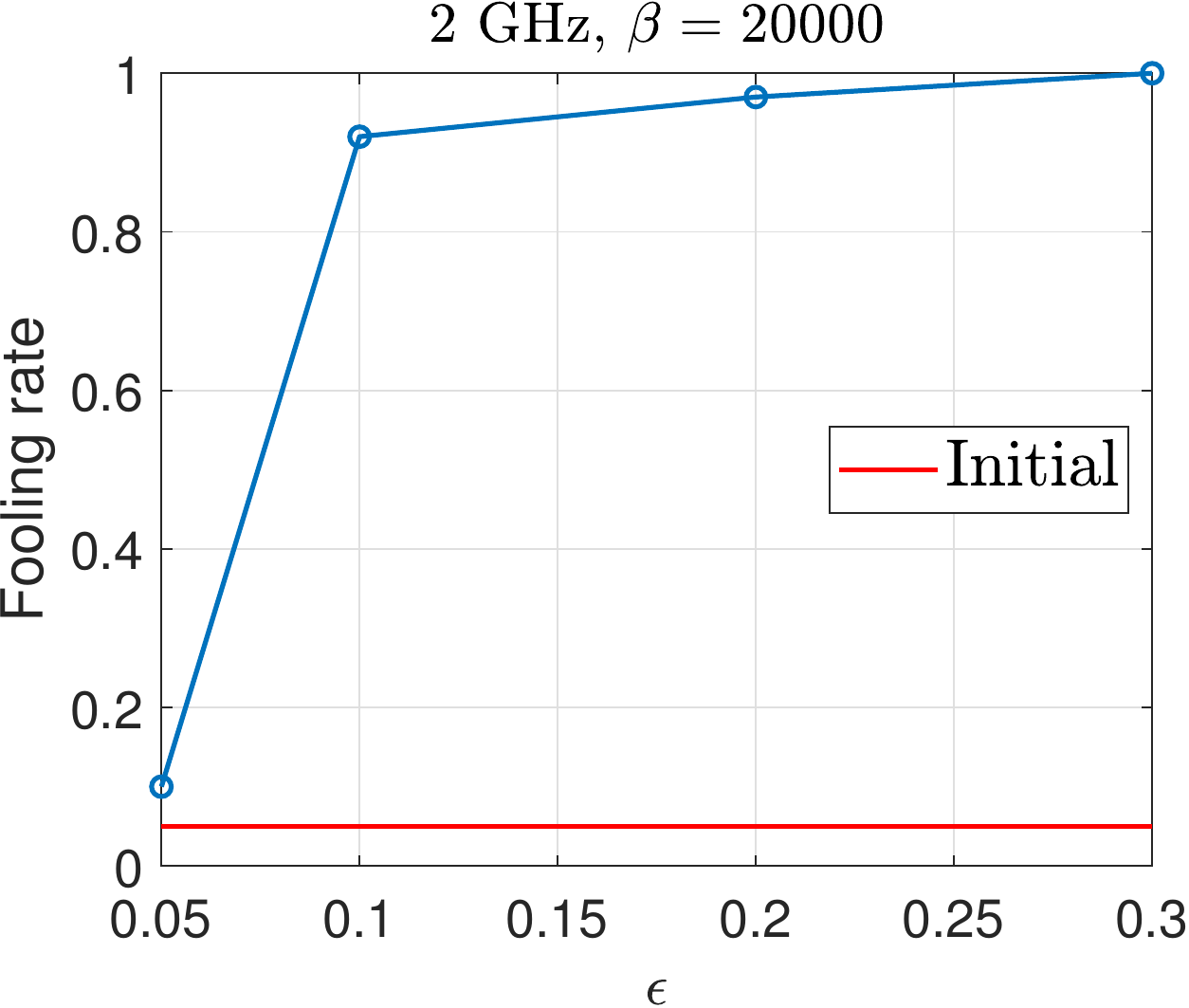}}
    \caption{{Experiment conducted to evaluate the effect of changing $\epsilon$ on the fooling rate. The testbed consisted of 8 ADALM Pluto SDRs.}}%
    \label{fig:fr_sdr}%
\end{figure}

We seek to understand a fundamental property of our algorithm in Experiment 3: is $G$ learning adversarial noise, or is it somehow learning to replicate the fingerprint of one of the transmitters in $\mT$? Unlike images where we may answer this problem with a visual inspection, we try to find an answer to this question numerically with the following experiment: first $G$ is allowed to converge on a particular instance of $D$, and the signals coming through that $G$ are tested on several other realizations of $D$---of different neural network architectures---trained to discriminate the same set of authorized transmitters $\mT$. If $G$ had actually learned to replicate RF fingerprints, it should achieve similar fooling rates irrespective of the particular $D$ it is being tested upon. To test our hypothesis, we trained 6 different realizations of $D$, $\mathcal{D}=\{\text{disc}_1, \text{disc}_2, \text{dclass}_1, \text{dclass}_2, \text{ova}_1, \text{ova}_2\}$; three different architectures \textit{disc}, \textit{dclass} and \textit{ova} were used and two instances each were created from each architecture. \textit{disc} is the binary discriminator architecture described in Section \ref{sec:nn_arch}. \textit{dclass} and \textit{ova} are two additional architectures defined and tested in \cite{hanna_spawc_2020} for RF fingerprinting, both sharing the same feature extractor given in Fig. \ref{fig:discriminator_classifier} and only differing in the classifier blocks used. \textit{dclass} consists of  a multi-class classifier having $|\mT|+1$ outputs; the first $|\mT|$ outputs correspond to the authorized transmitters and the last one corresponds to outliers. \textit{ova} has a single feature extractor shared across $|\mT|$ copies of the binary classifier block in Fig. \ref{fig:discriminator_classifier}, with the $i$-th such block denoting whether the signal is from the $i$-th transmitter or not. Table \ref{tab:fr_table} denotes the fooling rates observed when $G$ was allowed to attack $D=\text{disc}_1$ and signals from $T_A$  were tested on each discriminator in $\mathcal{D}$. This was then repeated with $D=\text{disc}_2$. It is clear that while a $G$ trained on either $\text{disc}_1$ or $\text{disc}_2$ could be used to attack the other with practically the same level of effectiveness, the effectiveness drops significantly when used against other architectures (albeit a significant increase in fooling rate). This confirms the hypothesis we set out to test; that is, $G$ learns to produce adversarial examples and does not learn actual RF fingerprints in $\mT$.

\begin{table}[H]
\setlength{\arrayrulewidth}{0.7pt}
\begin{tabular}{|c|c|c|c|c|c|c|}
\hline
\multirowcell{2}{$G$ is train- \\ed on} & \multicolumn{6}{c|}{$G$ is tested on} \\ \cline{2-7} 
                                 & $\text{disc}_1$ & $\text{disc}_2$ & $\text{dclass}_1$ & $\text{dclass}_2$ & $\text{ova}_1$ & $\text{ova}_2$    \\ \hline
$\text{disc}_1$                           &      \cellcolor{green!25}0.999 &	\cellcolor{green!25}0.999 &	\cellcolor{yellow!25}0.514 & \cellcolor{yellow!25}0.552 & \cellcolor{yellow!25}0.548 &	\cellcolor{red!25}0.449 \\ \hline
$\text{disc}_2$                           &    \cellcolor{green!25}1 &	\cellcolor{green!25}1 &	\cellcolor{red!25}0.276	& \cellcolor{yellow!25}0.893 &	\cellcolor{red!25}0.217 &	\cellcolor{red!25}0.359  \\ \hline
\end{tabular}
\caption{Effectiveness of a $G$ trained on $D=\text{disc}_1$ and $D=\text{disc}_2$ on some $D$s of different architectures}
\label{tab:fr_table}
\end{table}

For Experiment 4,  a binary discriminator was trained offline from a dataset captured on the SDR testbed; each transmitter took turns repeatedly transmitting the same predefined sequence of 256 IQ samples to the receiver, and the signals received at $R$ were collected for each transmitter. Then the impersonator started transmitting (authorized transmitters were inactive), and it was allowed to modify its  IQ samples before transmission according to Algorithm \ref{alg:cooperative}, using the feedback from the receiver. Note that the SDRs were simply used for over-the-air transmission and reception---all other operations such as training the attacker and calculating authentication decisions were done inside the computer. The results obtained, given in Fig. \ref{fig:fr_sdr_results}, closely resembles the trend suggested in the moderate SNR region in Fig. \ref{fig:snr_eps}. This means that our experimental results are consistent with the simulation results. 

\section{Conclusion and Future Work}
In this paper, we evaluated the feasibility of using policy gradient methods to penetrate a physical layer wireless authentication system which uses a passive deep-learning based classifier. We introduced an algorithm that adds carefully learned perturbations to the IQ samples transmitted by an adversarial transmitter to fool the authenticator into classifying it as an authorized transmitter. Experiments on a simulated wireless environment and an SDR testbed revealed that it is possible to fool the authenticator at extremely high fooling rates, using surprisingly little information---namely, a binary feedback from the authenticator indicating its decision and the modulation and pulse shaping used by the authorized transmitters. We also  showed that by limiting $\epsilon$, the distortion level of the impersonator signals could be kept low while still reaching a high fooling rate. Furthermore, we provided empirical evidence that our approach in fact produces adversarial examples and does not replicate the RF fingerprints of the transmitters.

While we only considered untargeted attacks, the possibility of launching targeted attacks with this method---where we try to impersonate a particulay transmitter in $\mT$---still remains. In a future work, we expect to present an algorithm for the case when $R$ is non-cooperative (it does not provide feedback endlessly), where an adversarial receiver is used instead to aid the impersonator. We also wish to evaluate the possible defenses that can be put in place against these types of attacks, both proactively and reactively.

\section*{Acknowledgments}

We wish to thank Samer Hanna (UCLA) for help in implementing the wireless system of the simulation environment.

\bibliographystyle{ieeetr}

\end{document}